\documentclass[psamsfonts,reqno]{amsart}

\usepackage{amssymb,amsfonts}
\usepackage[all,arc]{xy}
\usepackage{enumerate}
\usepackage{mathrsfs}
\usepackage{xypic}

\usepackage{amsmath,amssymb,amsfonts,wasysym}   
\usepackage{graphics}                           
\usepackage{amsmath}  
\usepackage{mathtools} 
\usepackage{hyperref}                           
\usepackage{comment}
\usepackage[english]{babel}

\theoremstyle{definition}

\theoremstyle{remark}

\makeatletter
\let\c@equation\c@thm
\makeatother
\numberwithin{equation}{section}

\title{Considerations on the genetic equilibrium law}

\author{Simone Camosso}

\date{}

\begin{document}

\begin{abstract}
In the first part of the paper I will present a brief review on the Hardy--Weinberg equilibrium and it's formulation in projective algebraic geometry. In the second and last part I will discuss examples and generalizations on the topic.

\end{abstract}

\maketitle

\tableofcontents

\section{Introduction}
The study of population genetics, evolution and its evolutionary trees are classical subjects in biology. A mathematical approach consists in the maximum likelihood estimation, a technique largely used in statistic. This approach leads to the problem of maximizing particular functions of certain parameters. A theoretical study and an upper bound for the maximum likelihood degree is discussed in \cite{CHKS}. Different techniques in order to solve likelihood equations are described in \cite{HKS}. Applications of these methods have been used when the statistical model is an algebraic variety, this is the case of Fermat hypersurfaces treated in \cite{AAGL}. In the other side we have applications of these ideas to biological models. In this direction we refer to a work of \cite{CKS} where phylogenetic models in two different topologies have been studied by the authors. 

The purpose of this paper is a ``soft'' introduction to these ideas with a discussion on the Hardy--Weinberg case.

\section{The Hardy--Weinberg law}

The Hardy--Weinberg law states that allele and genotype frequencies in a population remain constant during the generation change. This happens under the following assumptions: the size of the population must be very large, we have absence of migration and mutations, the mating is random and the natural selection doesn't affect the alleles under consideration. Mathematically if $p$ represents the number or pure dominants characters $AA$, $q$ the number of heterozygotes $Aa$ and $r$ the number of pure recessives $aa$, the following proportion holds $p:2q:r$ (see \cite{H}). Another way, if $p$ and $q$ represent the allele frequencies of the character $A$ and $a$ with $p+q=1$, taking the square we find that:

\begin{equation}
\label{HW1}
p^2+2pq+q^2\,=\,1,
\end{equation}
\noindent
where $p^{2},2pq$ and $q^{2}$ represent the genotype frequencies associated to $AA,Aa$ and $aa$. The equation $(\ref{HW1})$ describes the constancy of the genotypic composition of the population and is called the Hardy--Weinberg principle or the Hardy--Weinberg equilibrium (HWE). We consider \cite{FBB} and \cite{HC} as scholarly references on this subject. Different generalizations of $(\ref{HW1})$ are possible. The first concern the number of alleles at a locus. For example in the case of three alleles $A_{1},A_{2}$ and $A_{3}$, with frequencies respectively given by $p,q$ and $r$, the genotype frequencies are given by the following expansion $(p+q+r)^2\,=\,p^{2}+q^{2}+r^{2}+2pq+2qr+2pr$. In general, for any number $n$ of alleles with frequencies $p_{i}$, we have that:

\begin{equation*}
\label{HW2}
\left(p_{1}+ \ldots +p_{n}\right)^{2}\,=\,\sum_{i=1}^{n}p_{i}^{2}+\sum_{i\not=j}2p_{i}p_{j}.
\end{equation*}

In another direction the generalization is given considering the binomial $(p+q)^{m}$, with $m=3,4,5,\ldots$. This is the case of polyploid.  For example considering tetraploids ($m=4$) the procedure involves the expansion of $(p+q)^{4}$. We observe how in this particular example the frequency of heterozygotes (given by the mixed terms in the expansion) is $2pq(2-pq)$ that is considerably greater then $2pq$, the frequency for a diploid organism. More information on this topic can be found in \cite{FBB}.

\section{Projective and algebraic geometry}

In this section we shall examine how translate previous considerations in the modern language of projective and algebraic geometry. The setting is the same of \cite{HS} in its first lecture. We shall show how the Hardy--Weinberg law can be formulated in a fixed system of homogeneous coordinates $l,m,n$ in $\mathbb{P}^{2}$. First we shall consider the open triangle $\Delta_{2}=\{(l,m,n)\in\mathbb{R}_{+}^{3}: l+m+n\,=\,1\}$, where $\mathbb{R}_{+}$ are the positive reals. Second we shall observe that setting $l=p^{2},m=2pq,n=q^2$ to be genotype frequencies, we have the relation:

\begin{equation}
\label{parabola}
m^{2}\,=\,4ln,
\end{equation}
\noindent
that is the equation of a parabola in the triangle of vertex $(1,0,0),(0,1,0),(0,0,1)$. We call the zero locus of $(\ref{parabola})$, denoted also by $V(m^{2}-4ln)$, the Hardy--Weinberg curve (details are in \cite{E}). In the theory of \cite{HS} (and \cite{M}) is of particular interest a function called ``likelihood function'' that depends by some positive integer parameters. This function is positive on $\Delta_{2}$ and zero on the boundary of $\Delta_{2}$. We shall denote this function by $l$ and with $u_{0},u_{1},\ldots$ the corresponding parameters. In the case of the Hardy--Weinberg curve this function has the following form:

\begin{equation*}
\label{HW3}
l_{u_{0},u_{1},u_{2}}\,=\,l^{u_{0}}m^{u_{1}}n^{u_{2}}\,=\, 2^{u_{1}}p^{2u_{0}+u_{1}}q^{u_{1}+2u_{2}}.
\end{equation*}

We observe that $l_{u_{0},u_{1},u_{2}}$ is a function depending only by the variable $p$ (because $q=1-p$) and the MLE problem consists in the estimation of $p$ maximizing the function $l_{u_{0},u_{1},u_{2}}$. Lagrange Multipliers can be used in order to solve the problem and in this case the solution is given by the point:

\begin{equation}
\label{point}
\widehat{p}=\frac{2u_{0}+u_{1}}{2u_{0}+2u_{1}+2u_{2}}.
\end{equation}

\section{Examples, generalizations and conclusion}

As exercise we shall apply the same procedure in order to solve the MLE problem for the case of three alleles and in the second time for the case of tetraploids. For the first we shall start writing the ``likelihood function'' associated to the HWE given by $(p+q+r)^{2}$, where $p,q$ and $r$ are the usual frequencies. In this case we have that: 

\begin{equation*}
\label{HW4}
l_{u_{0},u_{1},u_{2},u_{3},u_{4},u_{5}}\,=\,2^{u_{3}+u_{4}+u_{5}}p^{2u_{0}+u_{3}+u_{5}}q^{2u_{1}+u_{3}+u_{4}}(1-p-q)^{2u_{2}+u_{4}+u_{5}}.
\end{equation*}

We shall proceed maximizing the function of two variables $p,q$. This is an ordinary problem of calculus that gives as answer the point:

\begin{equation*}
\label{HW5}
\left(\frac{2u_{0}+u_{3}+u_{5}}{2u_{1}+2u_{2}+2u_{4}+u_{3}+u_{5}},\frac{u_{4}-u_{5}+2u_{1}-2u_{0}}{2u_{1}+2u_{2}+2u_{4}+u_{3}+u_{5}},\frac{2u_{2}+u_{4}+u_{5}}{2u_{1}+2u_{2}+2u_{4}+u_{3}+u_{5}}\right).
\end{equation*}

For the tetraploid case, before to proceed, we shall observe that calling $l_{0}=p^4,l_{1}=q^4,l_{2}=4pq^3,l_{3}=4p^3q,l_{4}=6p^2q^2$ the genotype frequencies, the Hardy--Weinberg equilibrium can be represented by the following relation:

\begin{equation*}
\label{HW6}
l_{4}^{4}\,=\,\frac{1}{81}l_{0}l_{1}l_{2}l_{3}.
\end{equation*}

The associated ``likelihood function'' is

\begin{equation*}
\label{HW7}
l_{u_{0},u_{1},u_{2},u_{3},u_{4}}\,=\,l_{0}^{u_{0}}l_{1}^{u_{1}}l_{2}^{u_{2}}l_{3}^{u_{3}}l_{4}^{u_{4}}.
\end{equation*}

We shall make the expedient of consider the logarithm of the previous function instead the original finding as maximizing point:

\begin{equation*}
\label{HW8}
\left(\frac{u_{0}}{|u|},\frac{u_{1}}{|u|},\frac{u_{2}}{|u|},\frac{u_{3}}{|u|},\frac{u_{4}}{|u|}\right),
\end{equation*}
\noindent
where $|u|=u_{0}+u_{1}+u_{2}+u_{3}+u_{4}$. We recommended the use of a scientific software, as Maple or MATLAB, especially when the number of parameters is considerably high.  

Now I want to spend these last words comparing analogies between the HWE and the algebraic geometry. It is clear that a possible extension of the Hardy--Weinberg law can take the following form:

\begin{equation}
\label{HW9}
\left(p_{0}+\ldots+p_{n}\right)^{m}\,=\,c,
\end{equation}
\noindent
where $c$ is some constant and $p_{i}$ from $i=0,\ldots,n$ are the allele frequencies such that the sum is fixed. Now expanding $(\ref{HW9})$ we find the polynomial form:

$$\sum_{i_{0}+\ldots +i_{n}\,=\,m}\frac{m!}{i_{0}!\cdots i_{n}!}p_{0}^{i_{0}}\cdots p_{n}^{i_{n}}\,=\,c.$$ 

From the algebraic geometry point of view this is the image of the Veronese map $\nu_{m}:\mathbb{P}^{n}\rightarrow \mathbb{P}^{\binom{n+m}{m}-1}$ given by $(p_{0},\ldots,p_{n})\mapsto \left(p_{0}^{m},p_{0}^{m-1}p_{1},\ldots,p_{n}^{m}\right)$ (see \cite{A}). The classical Hardy--Weinberg law corresponds to the case of $\nu_{1}:\mathbb{P}^{1}\rightarrow \mathbb{P}^{2}$ that $(p,q)\mapsto (p^{2},pq,q^{2})$ and $c=1$. Using the identification between homogeneous polynomials that are power of linear forms and the image of the Veronese map, we can think these generalized laws as Veronese projective varieties. From the side of algebraic geometry there are a rich collection of results concerning the Veronese and Segre varieties, for example it is possible to compute the Hilbert polynomial and other invariants. It is not all peace and light because the constraint $\Delta_{n}=\{(p_{0},\ldots,p_{n})\in\mathbb{R}_{+}^{n+1}:p_{0}+\cdots +p_{n}=1\}$ doesn't permit the complete translation of the problem using the previous 
identification.

Anyway the methods of numerical algebraic geometry seem to give good prospects in this direction and in \cite{HRS} the ML degree has been calculated for matrices with rank constraints. In particular the case of rank one gives the ML degree equal to one, so $\widehat{p}$ is a rational function of a set of parameters $u_{0},u_{1},\ldots$.


\begin{thebibliography}{9}
\bibitem{AAGL}
D.Agostini, D.Alberelli, F.Grande, P.Lella, ``The maximum likelihood degree of Fermat hypersurfaces'', arXiv:1404.5745. 

\bibitem{A}E.Arrondo, ``Introduction to projective varieties'', unpublished notes from the website: http://www.mat.ucm.es/~arrondo/projvar.pdf (2007), 9--10.

\bibitem{CHKS}
 F.Catanese, S.Hoşten, A.Khetan, B.Sturmfels, ``The maximum likelihood degree'', Amer. J. Math. 128 (2006), no. 3, 671--697. MR 2230921.
 
 \bibitem{CKS} B.Chor, A.Khetan, S.Snir, ``Maximum likelihood on four taxa phylogenetic trees: analytic solutions'', The 7th Annual Conference on Research in Computational Molecular Biology--RECOMB 2003, Berlin, April 2003, pp. 76--83.
 
\bibitem{E}
A.W.F.Edwards, ``Foundations of Mathematical Genetics'', Cambridge University Press, Cambridge (2000).

\bibitem{FBB} 
R.Frankham, J.D.Ballou, D.A.Briscoe, ``Introduction to conservation genetics'', Cambridge (2002), 86--90.

\bibitem{H} 
G.H.Hardy, ``Mendelian Proportions in a Mixed Population'', Science, New Series, Vol.28, 706 (1908), 49--50.

\bibitem{HC} 
D.L.Hartl, A.G.Clark, ``Principles of population genetics'', Sinauer Associates, Inc. Publishers, Sunderland, Massachusetts (Fourth Edition).

\bibitem{HKS} S.Ho\c{s}ten, A.Khetan, B.Sturmfels, ``Solving the Likelihood Equations'', Foundations of Computational Mathematics, Vol. 5, Issue 4, pp 389--407, 2005.

\bibitem{HRS} J.Hauenstein, J.Rodriguez, B.Sturmfels, ``Maximum Likelihood for Matrices with Rank Constraints'', Journal of Algebraic Statistics, Vol.5, Issue 1 (2014), pp 18--38.

\bibitem{HS} 
J.Huh, B.Sturmfels, ``Likelihood Geometry'', Combinatorial Algebraic Geometry: Levico Terme, Italy 2013, Springer International Publishing, Vol.2108 of the series Lecture Notes in Mathematics (2014), 63--117.

\bibitem{M}
I.J.Myung, ``Tutorial on maximum likelihood estimation'', Journal of Mathematical Psychology 47 (2003) 90--100.
\end{thebibliography}
\end{document}